\documentclass[aps,prr,showpacs,superscriptaddress,twocolumn,raggedfooter,raggedbottom,nofootinbib]{revtex4-2}

\usepackage{graphicx}
\usepackage{dcolumn}
\usepackage{bm}
\usepackage{amsmath,amssymb}
\usepackage{amsthm}

\usepackage{hyperref}

\begin{document}

\preprint{APS/123-QED}

\title{On the mutual exclusiveness of time and position in quantum physics and the corresponding uncertainty relation for free falling particles}

\author{Mathieu Beau}
\affiliation{Physics department, University of Massachusetts, Boston, Massachusetts 02125, USA}

\author{Lionel Martellini}

\affiliation{Finance department, EDHEC Business School, Nice 06200, France}

\date{\today}

\begin{abstract}

The uncertainty principle is one of the characteristic properties of quantum theory, where it signals the incompatibility of two types of measurements. In this paper, we argue that measurements of time-of-arrival $T_x$ at position $x$ and position $X_t$ at time $t$ are mutually exclusive for a quantum system, each providing complementary information about the state of that system. For a quantum particle of mass $m$ falling in a uniform gravitational field $g$, we show that the corresponding uncertainty relation can be expressed as
$
\Delta T_x \Delta X_t \geq \frac{\hbar}{2mg}.
$
This uncertainty relationship can be taken as evidence of the presence of a form of epistemic incompatibility in the sense that preparing the initial state of the system so as to decrease the measured position uncertainty will lead to an increase in the measured
time-of-arrival uncertainty. These findings can be empirically tested in the context of ongoing or forthcoming experiments on measurements of time-of-arrival for free-falling quantum particles. 

\end{abstract}

\maketitle

\pagebreak

\section{Introduction}

While the Born rule gives the probability distribution of a position measurement at a fixed time, there is no readily available rule in the standard formalism of quantum mechanics for obtaining the probability distribution of a time measurement at a fixed position. This so-called \textit{time-of-arrival (TOA) problem} has been extensively debated in the literature, where a variety of competing approaches have been proposed (see \cite{muga2007time} for a review), but no consensus has emerged so far. The lack of an accepted formalism for the analysis of time-of-arrival can be regarded as one key blind spot in our quantum theoretical description of physical phenomena. In particular, it has been identified as one outstanding difficulty in the formulation of a coherent theory of quantum gravity (\cite{Anderson12} and \cite{Anderson17}). More pragmatically, it is also problematic in the context of interpreting empirical results related to free-falling quantum atoms, which is a prolific area of experimental research. Starting in the nineties, experimental techniques involving cold atoms have been developed to generate empirical TOA distributions using temporal slits (\cite{Dalibard95,Dalibard96,Dalibard96_2}). Recent technological advances have significantly enhanced the precision in the time measurements for free-falling objects, as demonstrated in projects such as MICROSCOPE \cite{Microscope17,Microscope22}, LISA-Pathfinder \cite{LISA18,LISA19}, free-falling matter waves \cite{QuantumTest14}, microgravity experiments on Earth \cite{MicrogravityEarth10,MicrogravityEarth13}, QUANTUS-MAIUS \cite{MicrogravityEarth13,MAIUS18}, and the Bose-Einstein Condensate and Cold Atom Laboratory (BECCAL) \cite{CAL21,CAL22,CAL23}, and it is expected that future experiments, including the Gravitational Behaviour of Anti-hydrogen at Rest (GBAR) experiment \cite{GBAR14,GBAR19,GBAR22,GBAR22_2} and the Space-Time Explorer and Quantum Equivalence Principle Space Test (STE-QUEST) \cite{Altschul15} could lead to further enhancements and yield new experimental insights into the analysis of time in quantum physics (see \cite{ExpReview21} for an extensive review of these developments). 


A particularly striking consequence of our incomplete understanding of the TOA problem is the fact that these empirical results cannot be compared with predictions for the mean and uncertainty of time-of-arrival distributions, which are not available for free-falling particles, or in fact for any quantum system. Some progress in this direction has fortunately been made in a recent paper focusing on time measurements for free-falling Gaussian systems (\cite{Beau24}), where standard results from statistics are used to derive within the standard formalism the exact expression for the TOA distribution directly from the Born rule. The straightforward stochastic representation introduced in \cite{Beau24} can also be used to derive approximate expressions for the mean value and the standard deviation of the TOA of the particle in the semi-classical and in the long time-of-flight (TOF) regime. In the quantum regime where the de Broglie wavelength becomes large compared to the initial spread of the wave packet, one striking finding, which directly follows from a Jensen-type inequality, is that the time of arrival (TOA) of a free-falling quantum particle at a given position is greater than the corresponding classical time-of-arrival, with a delay that depends on the mass of the particle.

The present article further contributes to the understanding of time-of-arrival measurements for free-falling particles by showing that an inverse relationship exists between the uncertainty over the measured position at a given time and the uncertainty over the measured arrival time at a given position. More precisely, we find that the product of measurement uncertainty over time-of-arrival $T_x$ at position $x$ and position $X_t$ at time $t$ for a quantum particle of mass $m$ falling in a uniform gravitational field $g$ is given by the following relation 
$
\Delta T_x \Delta X_t \geq \frac{\hbar}{2mg}.
$
More generally, and extending the discussion beyond free-falling quantum systems, we argue that this novel uncertainty relation signals the presence of a form of epistemic incompatibility between time and position measurements. We also provide a conceptual justification for the mutual exclusiveness of position and time measurements by carefully analyzing the types of stylized experimental protocols required to perform such measurements and showing that the presence of fundamental differences in their nature makes it intrinsically impossible to increase the precision in both.

\section*{Mutual exclusiveness of time and position in quantum physics}

\begin{figure*}
    \centering
    \includegraphics[scale=0.3]{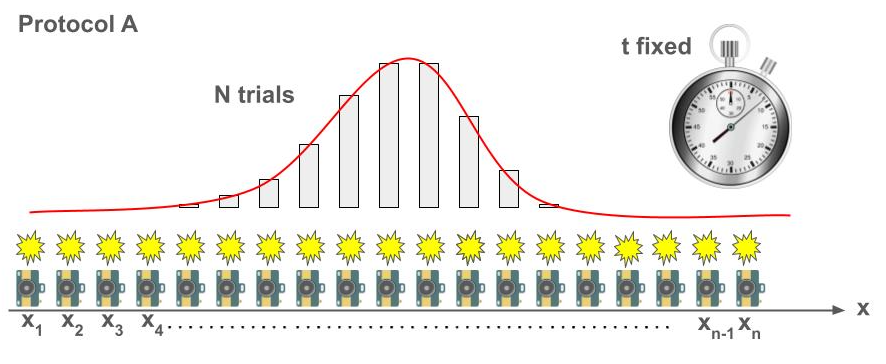}\includegraphics[scale=0.3]{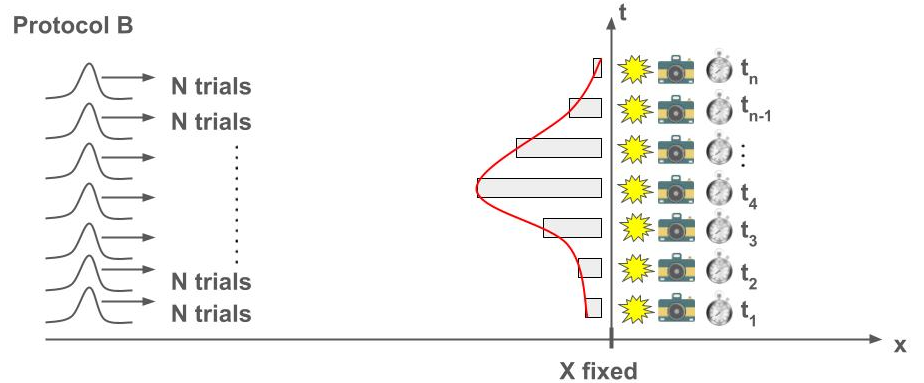}
    \caption{\textbf{Schematic representation of the two protocols A and B to measure the position-distribution and time of arrival-distribution.} On the left panel, we represent the position distribution that is constructed from the measurement of the position of a particle using detectors located at positions $x_k,\ k=1,\cdots,n$ switched on at time $t$. We repeat the same experiment $N$ times and record each time the location where the particle is detected. On the contrary, the protocol B shown on the right panel, which allows for the empirical construction of the time-of-arrival distribution, consists of measuring the presence of a particle with one single detector located at a fixed position $x$ that is switched on at a given time $t_1$. If the particle is detected, we record it and repeat the experiment $N$ times to increase the power of the associated statistical measure. We then repeat the same experiment but choose a different time of detection $t_2$, then $t_3$, then $t_4$, ..., $t_n$. Notice that repeated measurements should not be allowed with the detector being switched on at $t=t_1,\ t_2,\ ..,\ t_n$ for the same experiment, which explains why we need a larger total number of trials compared to protocol A. Repeated measurement would indeed destroy the quantum coherence of the particle, and would not allow to construct an empirical distribution for the time of a first measurement at position $x$.} 
    \label{Fig:Protocols}
\end{figure*}

In classical physics, one can measure both the position and time-of-arrival of a moving object within one single unified experimental setup. For this, one may use an external clock to measure time-of-arrival, and detectors placed at various points in space (or alternatively a high-resolution camera) to measure position. Each position measurement corresponds to a point in the space-time representation of the trajectory. By detecting the object position and recording the time of measurement, we can thus represent the object position as a function of time $x=f(t)$. Even though the precision of these measurements is limited by the resolution of the detector and the resolution of the clock, one can at least in principle engineer measurement devices precise enough to ensure that the idealized model $x=f(t)$ provides a satisfactory description of the physics of the underlying system. 

The situation is radically different in quantum physics, where it turns out that position measurements and TOA measurements cannot be performed within a single coherent experimental setup. There indeed exist two different, and mutually exclusive, experimental protocols involved in measurements of the position at a given time $t$ and the time-of-arrival at a given position $x$. As a result, the epistemic nature of the measurement of position and time-of-arrival differ significantly at the experimental level. 
To see this, let us examine the following two experimental protocols. In the first experiment A, the objective is to measure the particle position at a specific time, while in the second experiment B, the aim is to measure the time-of-arrival (TOA) of the particle at a given position, see Figure \ref{Fig:Protocols}. Protocol A consists of (i) first placing $n$ synchronized detectors at different positions $x_1,\ x_2,\ \cdots,\ x_{n}$ with a spatial resolution $\delta x$, (ii) then recording the position of the particle $X_t$ at the time $t$ (here $X_t$ denotes a stochastic variable that represents the outcome of the position measurement at time $t$ experiment), and (iii) finally repeating steps (i) and (ii) a large number $N$ of times to construct the probability distribution $\rho_t(x)$ at a fixed time $t$ \cite{Beau24}. 
As for the second experiment B, aiming to measure the time-of-arrival (TOA) at a given position $x$, we consider the following procedure (see Figure \ref{Fig:Protocols} and \cite{Beau24}). In the first step (i), we drop a particle at time $t=0$ and switch on a detector located at a fixed position $x$ at a later time $t_1$ and then (ii) repeat the same procedure $N$ times and record the number of times the particle has been detected at the same time $t_1$. As a final step (iii), we then redo steps (i) and (ii) after changing the time at which the detector is switched on, i.e., we perform $N$ trials for $t=t_2=t_1+\delta t$, where $\delta t$ is the temporal resolution, $N$ more trials for $t=t_3=t_2+\delta t$, and so on, until the final $N$ measurement at $t=t_n=t_1+(n-1)\delta t$. At the end of the experiment, we can reconstruct the time distribution $\Pi_x(t)$ of the associated random variable $T_x$, which represents the stochastic time-of-arrival (TOA) at the fixed position $x$. 
One could argue that instead of using the protocol B that leads to measuring the position $x$ at each time $t_k,\ k=1,\cdots, n$, one might consider instead a more direct protocol B' by activating the detector located at the position $x$ at $t=0$ and leaving it on during the measurement time $t_1\leq t\leq t_n$. This experimental protocol B', which would be perfectly symmetrical to protocol A, would however involve \textit{repeated measurements}, implying continuous interaction between the detector and the particle for all $t\geq 0$, leading to changes in the state of the underlying system. While a review of the literature on repeated measurements is beyond the scope of this article (see \cite{davies1969quantum} and \cite{davies1970repeated} for early papers on the subject), we remark that such repeated interactions between the quantum system and the measuring device would lead to measuring a time would no longer correspond to the TOA defined in this article as the time of a \textit{first} measurement at position $x$.

From the discussion above, we conclude that A and B appear to be mutually exclusive experimental protocols that cannot be performed simultaneously, implying that position and TOA are epistemically incompatible (see \cite{Bohr28} and Chapter 5 in \cite{Murdoch87} for the analog definition in the case of position versus momentum). In what follows, we show that this mutual exclusiveness of position and TOA measurements is quantitatively associated with a novel type of uncertainty relation for which we provide an explicit expression for free-falling particles. While we specialize the analysis of free-falling systems in what follows, the above discussion about the epistemic incompatibility of time-of-arrival and position experimental measurements suggests that the associated uncertainty relation holds more generally, a question that we in the concluding section of this paper.

\section*{Uncertainty relation between time-of-arrival and position for a free-falling particle}

In what follows we use the stochastic representation recently introduced in \cite{Beau24} to analyze the relation between the random variable, denoted by $T_x$, that describes a time-of-arrival measurement at a given position $x$, and the random variable, denoted by $X_t$, that describes a position measurement at a given time $t$. As in \cite{Beau24}, we focus the analysis on a free-falling quantum particle in 1 dimension, and we assume an initial Gaussian state. For the free-falling particle (and also for the free particle, the simple and time-dependent harmonic oscillator, constant or time-dependent electric fields), it can be shown that the system stays Gaussian at all times when evolving from a Gaussian initial state (see for example \cite{Klebert73}). 

As a first step, we note that $X_t$ can be written with no loss of generality in the Gaussian setting as \cite{Beau24}:
\begin{equation}\label{Eq:rv:gaussian:t>0}
    X_t  = x_c(t) + \xi\sigma(t),\ 
\end{equation}
where $\xi=\mathcal{N}(0,1)$ is a normally distributed random variable with a variance of $1$ and a mean value of $0$, and where $\sigma(t)$ is the standard-deviation of the Gaussian distribution that is centered at the classical path $x_c(t)$, which by the correspondence principle is also the mean value of the position operator $\langle \hat{x}_t\rangle = \int_{-\infty}^{+\infty}x\rho_t(x)dx = x_c(t)$).

Assuming a free-falling particle with a zero-mean position and restricting the analysis to a dropped particle with a zero-mean initial velocity, we can write the classical position at time $t$ as $ x_c(t) = \frac{g}{2}t^2$ and the standard deviation of the position at time $t$ as $\sigma(t) = \sqrt{1+\frac{t^2}{\tau^2}}$ with $\tau = \frac{2m\sigma^2}{\hbar}$. Since the distribution of $\xi$ is a normalized Gaussian $\frac{1}{\sqrt{2\pi}}\exp{\left(-\frac{\xi^2}{2}\right)}$, we confirm by a linear transformation that the distribution of the position is $\rho_t(x) = \frac{1}{\sqrt{2\pi \sigma(t)^2}}\exp{\left(-\frac{(x-x_c(t))^2}{2\sigma(t)^2}\right)}$, as expected. 

In a second step, and by definition of it being the first passage time at the position $x$, we note that the TOA $T_x$ can be written as $T_x = \inf\{t|X_t=x\}$. We then use $x = X_{T_x}$, where $X_t$ is given in
equation \eqref{Eq:rv:gaussian:t>0}, to obtain 
\begin{equation}\label{Eq:FF}
    x = \frac{g}{2}T_x^2 + \sigma \xi \sqrt{1+\frac{T_x^2}{\tau^2}}, \ 
\end{equation}
This quartic equation can be solved analytically to give the following explicit representation for the TOA $T_x$:
\begin{equation}\label{Eq:TxSol}
    T_x  = t_c \sqrt{1+2q^2\xi^2-2q\xi\sqrt{1+q^2\xi^2+\frac{1}{4q^2}\frac{\sigma^2}{x^2}}},\ \text{with}\ \xi\leq \frac{x}{\sigma}\ ,
\end{equation}
where $t_c =\sqrt{\frac{2x}{g}}$ is the classical time, where the factor $q =  \frac{\hbar}{2m\sigma\sqrt{2gx}} = \frac{\lambda}{4\pi\sigma}$ measures the ratio of the height-dependent de Broglie wavelength ($\lambda=\frac{\lambda}{\sqrt{2mE}}$ with $E=mgx$) to the initial width of the particle wave-packet,
and where the parameter $\beta \equiv \frac{1}{q}\frac{\sigma}{x}$ determines the distinction between the far-field ($\beta \ll 1$) versus near-field ($\beta \gg \max{(1,q)}$) regime of the system.\footnote{The near-field regime is obtained when the term $\beta^2=\frac{1}{4q^2}\frac{\sigma^2}{x^2}$ in equation \eqref{Eq:TxSol} dominates all the other terms, and specifically the term $1+q^2\xi^2$. If $q\xi\ll 1$, then $\beta^2\gg 1$. However, when $\xi^2$ is too large and $q\xi\gg 1$, $\beta^2$ must be substantially larger than that latter term. Since $\xi$ follows a normal distribution with a standard deviation of $1$, it suffices that $\xi$ does not become excessively large compared to $1$ 
Therefore, it is sufficient that $\beta \gg q$ and so that $\frac{\sigma}{x} \gg q^2$. In summary, we conclude that $\frac{\sigma}{x}$ must be significantly larger than the maximum value of $1$ and $q$.}

In what follows, we use \eqref{Eq:TxSol} to obtain the following uncertainty relation
\begin{equation}\label{Eq:Uncertainty:Farfield}
\Delta T_{x}\Delta X_{t}\geq \frac{\hbar }{2mg}, 
\end{equation} 
which holds for both the near-field and the far-field regimes. To see this, and starting with the far-field regime ($\frac{\sigma}{x} \ll q$), we first note that equation \eqref{Eq:TxSol} can be simplified as:
\begin{align}\label{Eq:TxFarField}
    T_x  &= t_c \left(\sqrt{1+q^2\xi^2}-q\xi\right) \nonumber \\ 
    &\approx 
    \begin{cases}
        t_c\left(1-q\xi + \frac{1}{2}q^2\xi^2\right),\ \text{if}\ q\ll 1 \\
        qt_c\left(|\xi|-\xi\right)\ \ \ \ \  \ \ \ \ \ ,\ \text{if}\ q\gg 1
    \end{cases}\ .
\end{align}
We then compute the uncertainty around the TOA measurement from these expressions. 
For $q\ll 1$, it was shown in \cite{Beau24} (see equation (17) with $v_0=0$) that $\Delta T_x = t_c\frac{\sigma}{\sqrt{2gx}\tau}$, whence $\Delta T_x = \frac{\hbar}{2mg\sigma}$, leading to the relation \eqref{Eq:Uncertainty:Farfield} as $\Delta X_t = \sigma(t) = \sigma\sqrt{1+\frac{t^2}{\tau^2}}\geq \sigma$. For $q\gg 1$, we thus find that $\Delta T_{x} = \frac{\hbar }{2mg\sigma}\sqrt{\frac{2(\pi-1)}{\pi}}$, which leads to
\begin{equation}\label{Eq:Uncertainty:Farfield:qlarge}
\Delta T_{x}\Delta X_{t}\geq \frac{\hbar }{2mg}\sqrt{\frac{2(\pi-1)}{\pi}} , 
\end{equation}
which is greater than $\frac{\hbar }{2mg}$ as $\sqrt{\frac{2(\pi-1)}{\pi}} >1 $.

In the near-field regime ($\frac{\sigma}{x} \gg \max(q,q^2)$), we obtain the following asymptotic relation:
\begin{equation}\label{Eq:TxNearField}
T_x \approx t_c \sqrt{1-\frac{\sigma}{x}\xi},\ \text{with}\ \xi\leq \frac{x}{\sigma}\ ,
\end{equation}
from where we can derive the approximations\footnote{Throughout the text, we use $\mathbb{E}(Y)$ to denote a mathematical expectation, as is customary for a random variable $Y$ (e.g., the TOA $T_x$ or the position $X_t$), while we use instead the notation $\langle\hat{A}\rangle$, as is customary for a quantum operator $\hat{A}$.} $\mathbb{E}(T_x)\approx \frac{2^{1/4}\Gamma\left(\frac{3}{4}\right)}{\sqrt{\pi}}t_c\sqrt{\frac{\sigma}{x}}$ and $\mathbb{E}(T_x^2)\approx \sqrt{\frac{2}{\pi}} t_c^2\frac{\sigma}{x}$ (see \cite{SM}), from which we obtain $\Delta T_x \approx k\cdot t_c \sqrt{\frac{\sigma}{x}} = k\cdot \sqrt{\frac{2\sigma}{g}} $, where $k\equiv \sqrt{\sqrt{\frac{2}{\pi}}\left(1-\frac{\Gamma\left(\frac{3}{4}\right)^2}{\sqrt{\pi}}\right)} \approx 0.349$. Finally, we have:
\begin{equation}\label{Eq:Uncertainty:Nearfield}
\Delta T_{x}\Delta X_{t} \geq k\cdot \sqrt{\frac{2\sigma^3}{g}} .
\end{equation}
This relation implies that as the initial uncertainty around the position $\sigma$ becomes very large, the lower bound of \eqref{Eq:Uncertainty:Nearfield} increases as $\sigma^{3/2}$. Since $\sigma$ must at least satisfy the condition $\frac{\sigma}{q x} \gg q$ for the asymptotic relation \eqref{Eq:Uncertainty:Nearfield} to be valid, we find that $\sigma \gg \frac{x_0}{2^{2/3}} \approx 0.630 x_0$, where $x_0 \equiv \left(\frac{\hbar^2}{2m^2g}\right)^{1/3}$ is the characteristic gravitational length \cite{Nesvizhevsky15}, which means that the lower bound in Equation \eqref{Eq:Uncertainty:Nearfield} is very large compared to $\frac{k}{\sqrt{2}}\frac{\hbar}{2mg} \approx 0.247 \frac{\hbar}{2mg}$ and thus very large compared to the lower bound in Equation \eqref{Eq:Uncertainty:Farfield}. In the intermediate regime $\frac{\sigma}{x} \sim \max(q,q^2)$ no analytical expression can be obtained for the lower bound of $\Delta T_x \Delta X_t $, but we confirm the uncertainty relation after a thorough exploration of the parameter space via numerical computation. Taken together, these results suggest that the bound in equation \eqref{Eq:Uncertainty:Farfield} is universal and is valid for all possible values of the parameters $g,\ x,\ m$, and $\sigma$. 

\begin{figure*}
    \centering
    \includegraphics[scale=0.5]{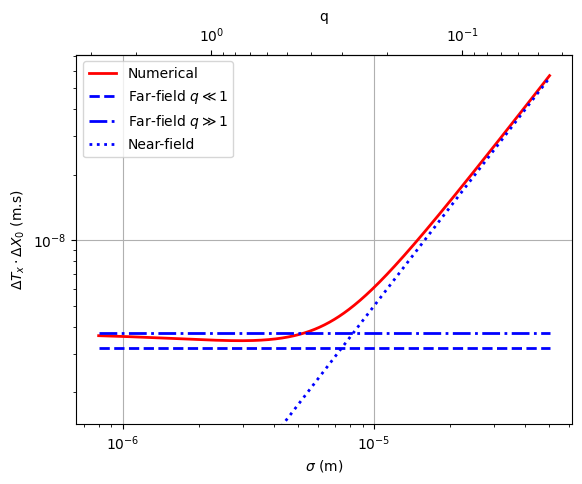}\hspace{0.5cm}\includegraphics[scale=0.5]{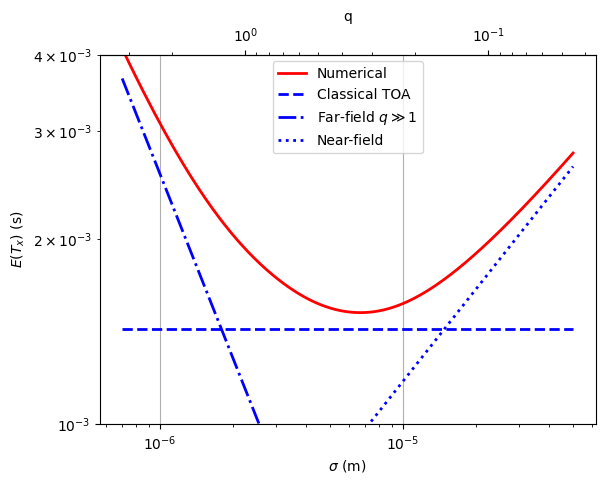}
    \caption{\textbf{TOA-position uncertainty relation and the mean value of TOA for a free-falling particle.} On the left panel, we show the value of the product $\Delta T_x \Delta X_0$ as a function of the initial standard deviation $\Delta X_0 = \sigma$  obtained from the numerical integration of the standard deviation of the stochastic variable given by equation \eqref{Eq:TxSol} (continuous-red-line), as well as the values obtained in the far-field regime as per equation \eqref{Eq:Uncertainty:Farfield} for $q\ll 1$ (dashed-blue-line) and equation \eqref{Eq:Uncertainty:Farfield:qlarge} for $q\gg 1$ (dashed-dotted-blue-line), and in the near-field regime \eqref{Eq:Uncertainty:Nearfield} (dotted-blue-line). As we see from this graph, the bound $\frac{\hbar}{2mg}$ (bottom dashed-blue line) is universal for all regimes. 
    On the right panel, we display the mean values of the TOA (continuous-red-line) as well as the classical TOA $t_c= \sqrt{\frac{2x}{g}}$ (dashed-blue-line) and the two asymptotes corresponding to the near-field (dotted-blue-line) and the far-field with $q\gg 1$ (dashed-dotted-blue-line). Notice that the mean value of the TOA is always greater than the classical TOA, showing the discrepancy between the two values of TOA obtained from experience A (classical TOA) and B (red-continuous curve).  In these two graphs, we took the values for $x=10^{-5} \text{m},\ g=9.8 \text{m}\cdot\text{s}^{-2}$ and $m=1.67 \cdot 10^{-27}\text{kg}$ (hydrogen atom). We added a secondary $x-$axis at the top to visualize the evolution of $q$ as a function of $\sigma$ as well.} 
    \label{fig:uncertainty}
\end{figure*}

In Figure \ref{fig:uncertainty}, we report the results of a numerical estimation of the evolution of $\Delta X_0\Delta T_x$ as a function of $10^{-2}\geq\frac{\sigma}{x}\geq 10^1$ for a hydrogen atom of mass $m=1.67\times 10^{-27} \text{kg}$ falling into the Earth's gravitational field $g=9.81 \text{m}\cdot\text{s}^{-2}$ with $x=10^{-6}\text{m}$. Figure \ref{fig:uncertainty} shows that $\frac{\hbar }{2mg}$ is indeed the lowest bound for all regimes, a result that is robust with respect to changes in parameter values.  \\

As a side comment, we note that the standard time/energy uncertainty relation can be confirmed within our framework. To see this, we first compute the uncertainty for the energy of a free-falling particle as $\Delta E = mg\sigma\sqrt{1+\frac{\hbar^4}{32g^2m^4\sigma^6}}$ \cite{SM}, and then combine this result with equation \eqref{Eq:Uncertainty:Farfield} to obtain as expected
\begin{equation}\label{Eq:TimeEnergyUncertainty}
    \Delta E \Delta T_x \geq \frac{\hbar}{2} .
\end{equation}
The time/energy uncertainty relation that we confirm here relates to a joint analysis of the dispersion of energy measurements for a quantum system and the dispersion of time-of-arrival measurements at a given position $x$. This interpretation of the time/energy uncertainty relation closely aligns with that presented in \cite{Dalibard96}, where the authors experimentally measure the dispersion in the time-of-arrival and compare it with the dispersion of the energy of the system. This is in contrast with competing interpretations of the time/energy relation, where time is rather regarded as the time of transition between different energy states (see \cite{Busch2008} for an extensive review). 
 
\section*{Discussion}

To interpret the uncertainty relation \eqref{Eq:Uncertainty:Farfield}, it is useful to revert to the discussion of the experimental protocols A and B used in position versus time measurements, as described in the first section of this paper. The time/position uncertainty relation introduced in this paper implies that if one repeats experiments A and B with initial states prepared with a smaller value of the initial wave packet dispersion $\sigma$, then the uncertainty for the position as measured through experiment A will decrease while the uncertainty for the time of arrival (TOA) as measured with experiment B will increase. 
Consequently, it is impossible to \textit{simultaneously} decrease both uncertainties to zero. Note that the term \textit{simultaneously} here refers to performing two experiments after preparing the initial state under the same conditions. This notion of simultaneity, which we call \textit{epistemic simultaneity}, is quite different from the standard causal notion of simultaneity defined in a space-time representation of physics
\footnote{Indeed, it does not make sense to envision measuring the time of arrival and the position \textit{simultaneously in time} since there is no single coherent experimental protocol that could be used to perform such a joint measurement, as discussed in the first section of this paper.}. 
Specifically, this notion arises when we assume that after preparing a state and performing some measurement from this state (e.g., position at a given time), and then replicating the same preparation and subsequently conducting a distinct measurement from this same state (e.g., time of arrival at a given position), we can statistically gain additional knowledge (or \textit{episteme}) into the nature of this state. The fact that the state is prepared identically (under the same experimental conditions) implies that two data samples obtained from the two different experiments, each with many trials, represent two \textit{simultaneous} pieces of information about the same system. This \textit{simultaneous acquisition} of sample data leads to complementary knowledge of the quantum system, and we thus conclude that (a) the measurements of position and time-of-arrival are not only \textit{mutually exclusive} in that they cannot be both measured simultaneously-in-the-temporal-sense, but also, and perhaps more importantly, that (b) they are \textit{epistemically incompatible} in that they cannot be both simultaneously-in-the-epistemic-sense measured with arbitrary precision.    

Bohr's definition of the \textit{epistemic incompatibility} of position and momentum (see \cite{Murdoch87}, Chapter 5) is that (a') the measurements are mutually exclusive and (b') the quantum of action $\hbar$ implies ``an indeterminable disturbance on the object'' after the measurement. According to Bohr, the combination of those two conditions implies the existence of the Heisenberg uncertainty relations (see \cite{Murdoch87}, Chapter 5) 
and defines the notion of position-momentum \textit{complementarity}. 
Bohr further posits that the complementarity of position and momentum is encoded in the commutation relation $[\hat{x},\hat{p}]=i\hbar$, implying that the measurement of the momentum first will disturb the movement of the object with an amplitude equal to $\hbar$, and thus prevents one from measuring the position and the momentum simultaneously in time. 
In the case of our study, Bohr's definition does not hold in the strict sense because (i) we did derive the time-position uncertainty relation without introducing a time-of-arrival operator, and (ii) we introduced a different notion of simultaneity based on the knowledge acquired from two different experiments performed under the same initial conditions. 
Despite some similarities with Bohr's definition of complementarity (mutual exclusiveness is common to both approaches), our notion of \textit{complementarity} for time-of-arrival and position is thus based on different criteria compared to momentum and position (criterion (b) versus (b') described above). Actually, it can be argued that our definition of complementarity extends Bohr's definition in the sense that it can be applied to any two mutually exclusive measurements regardless of whether or not these observables are associated with Hermitian operators. In this context, it also provides some potentially useful clarification into the time-energy uncertainty relation (see equation \eqref{Eq:TimeEnergyUncertainty} and related discussion). 

\section*{Conclusion}

In conclusion, this article demonstrates that measurements of time-of-arrival and position are mutually exclusive for a quantum particle, each providing complementary information about the system. For a free-falling particle, we also show that this mutual exclusiveness is associated with a novel uncertainty relation between TOA and position that can be experimentally tested in relevant platforms. It is worth mentioning that the process of determining optimal experimental conditions can be easily extended to the motion of charged particles in constant and uniform electric fields. 

From the experimental perspective, we also expect the uncertainty relation to have consequences in quantum metrology, in particular in measures of the acceleration due to gravity based on a detector for which the position is known, as is, for example, the case with the GBAR experiment \cite{GBAR14,GBAR19,GBAR22,GBAR22_2}. First, if the position-detector were located close to the source (near-field), then the analysis of the trajectories ought to be performed in the light of the results presented in this paper, in particular equations \eqref{Eq:TxNearField}-\eqref{Eq:Uncertainty:Nearfield} (as well as equations in the paragraph in between) and Figure \eqref{fig:uncertainty}. Secondly, the uncertainty relation \eqref{Eq:Uncertainty:Farfield} may have an impact on the parameter estimation error of $g$, which would lead to a modification of the Cramer–Rao bound used in \cite{GBAR22}. While beyond the scope of this paper, a modification of the Cramer–Rao bound and its quantum version using a stochastic representation of time-of-arrival would be a possibly relevant avenue for future research. Another potentially interesting metrological application would relate to enhancements of statistical procedures used to determine a detector's position $x$. After performing an experiment following protocol B, see Figure \ref{Fig:Protocols}, we can deduce the position from the value of the mean TOA $\mathbb{E}(T_x)$, which depends directly on $x$. There again the existence of the time/position uncertainty relation \eqref{Eq:Uncertainty:Farfield} (see also Figure \ref{fig:uncertainty}) implies the presence of intrinsic limitations to the precision of measurements of the detector position $x$ regardless of how small or large will be chosen the initial dispersion value $\sigma$. 

The mutual exclusiveness between TOA and position measurement, which has been shown in full generality in the first section of this article, suggests that a time/position uncertainty relation also holds for other systems, such as particles moving in time-dependent electric fields in time-dependent harmonic traps (which could potentially be realized in laboratory settings). At this stage, we conjecture the existence of a time-position uncertainty relation of the form $\Delta T_x \Delta X_t \geq qt_c\times \sigma = \frac{t_c}{\sqrt{2mE}}\frac{\hbar}{2}$, where $q = \frac{\lambda}{4\pi\sigma}$ indicates the degree of quantumness of the particle (increasing $q$ signaling increasing quantumness), $t_c$ is the classical time of arrival, and where $\lambda = \frac{h}{\sqrt{2mE}}$ is the de Broglie wavelength with $E$ being the classical energy of the particle. In the case analyzed in this paper, we obtain $E = mgx$ and $t_c=\sqrt{\frac{2x}{g}}$ for a free-falling particle released from rest, which consistently leads to $\Delta T_x \Delta X_t \geq \frac{\hbar}{2mg}$. In other words, we envision that the product of the standard deviations of TOA and position is bounded below by a semi-classical limit that depends on the quantum of action and classical quantities such as time and energy. As the quantumness factor $q$ increases, the lower bound of the uncertainty relation increases as well, as one would naturally expect. 
These time/position uncertainty relations could be explored for a large variety of quantum systems, both experimentally in laboratories and mathematically within the standard formalism of quantum mechanics. Indeed, the stochastic representation introduced in \cite{Beau24} and further developed in this article, provides a tool for analyzing the time-of-arrival distribution in various quantum systems. 
Of particular interest would be the analysis of discrete spin systems, where analogous uncertainty relations between transition time and spin states could be derived. We leave these, and other related questions, for further investigation. 

On a different note, it would also be useful to explore the interplay between time-position complementarity and space-time symmetry when quantum physics and relativity are intertwined. If the time/position uncertainty relation \eqref{Eq:Uncertainty:Farfield} introduced in this paper is experimentally confirmed, it must have a bearing within the non-relativistic limit of any quantum mechanics theory incorporating relativistic aspects.

\bibliography{Arxiv_v1_Beau}

\providecommand{\noopsort}[1]{}\providecommand{\singleletter}[1]{#1}%
\begin{thebibliography}{32}%
\makeatletter
\providecommand \@ifxundefined [1]{%
 \@ifx{#1\undefined}
}%
\providecommand \@ifnum [1]{%
 \ifnum #1\expandafter \@firstoftwo
 \else \expandafter \@secondoftwo
 \fi
}%
\providecommand \@ifx [1]{%
 \ifx #1\expandafter \@firstoftwo
 \else \expandafter \@secondoftwo
 \fi
}%
\providecommand \natexlab [1]{#1}%
\providecommand \enquote  [1]{``#1''}%
\providecommand \bibnamefont  [1]{#1}%
\providecommand \bibfnamefont [1]{#1}%
\providecommand \citenamefont [1]{#1}%
\providecommand \href@noop [0]{\@secondoftwo}%
\providecommand \href [0]{\begingroup \@sanitize@url \@href}%
\providecommand \@href[1]{\@@startlink{#1}\@@href}%
\providecommand \@@href[1]{\endgroup#1\@@endlink}%
\providecommand \@sanitize@url [0]{\catcode `\\12\catcode `\$12\catcode `\&12\catcode `\#12\catcode `\^12\catcode `\_12\catcode `\%12\relax}%
\providecommand \@@startlink[1]{}%
\providecommand \@@endlink[0]{}%
\providecommand \url  [0]{\begingroup\@sanitize@url \@url }%
\providecommand \@url [1]{\endgroup\@href {#1}{\urlprefix }}%
\providecommand \urlprefix  [0]{URL }%
\providecommand \Eprint [0]{\href }%
\providecommand \doibase [0]{https://doi.org/}%
\providecommand \selectlanguage [0]{\@gobble}%
\providecommand \bibinfo  [0]{\@secondoftwo}%
\providecommand \bibfield  [0]{\@secondoftwo}%
\providecommand \translation [1]{[#1]}%
\providecommand \BibitemOpen [0]{}%
\providecommand \bibitemStop [0]{}%
\providecommand \bibitemNoStop [0]{.\EOS\space}%
\providecommand \EOS [0]{\spacefactor3000\relax}%
\providecommand \BibitemShut  [1]{\csname bibitem#1\endcsname}%
\let\auto@bib@innerbib\@empty
\bibitem [{\citenamefont {Muga}\ \emph {et~al.}(2007)\citenamefont {Muga}, \citenamefont {Mayato},\ and\ \citenamefont {Egusquiza}}]{muga2007time}%
  \BibitemOpen
  \bibfield  {author} {\bibinfo {author} {\bibfnamefont {G.}~\bibnamefont {Muga}}, \bibinfo {author} {\bibfnamefont {R.~S.}\ \bibnamefont {Mayato}},\ and\ \bibinfo {author} {\bibfnamefont {I.}~\bibnamefont {Egusquiza}},\ }\href@noop {} {\emph {\bibinfo {title} {Time in quantum mechanics}}},\ Vol.\ \bibinfo {volume} {734}\ (\bibinfo  {publisher} {Springer Berlin, Heidelberg},\ \bibinfo {year} {2007})\BibitemShut {NoStop}%
\bibitem [{\citenamefont {Anderson}(2012)}]{Anderson12}%
  \BibitemOpen
  \bibfield  {author} {\bibinfo {author} {\bibfnamefont {E.}~\bibnamefont {Anderson}},\ }\bibfield  {title} {\bibinfo {title} {Problem of time in quantum gravity},\ }\href {https://doi.org/https://doi.org/10.1002/andp.201200147} {\bibfield  {journal} {\bibinfo  {journal} {Annalen der Physik}\ }\textbf {\bibinfo {volume} {524}},\ \bibinfo {pages} {757} (\bibinfo {year} {2012})},\ \Eprint {https://arxiv.org/abs/https://onlinelibrary.wiley.com/doi/pdf/10.1002/andp.201200147} {https://onlinelibrary.wiley.com/doi/pdf/10.1002/andp.201200147} \BibitemShut {NoStop}%
\bibitem [{\citenamefont {Anderson}(2017)}]{Anderson17}%
  \BibitemOpen
  \bibfield  {author} {\bibinfo {author} {\bibfnamefont {E.}~\bibnamefont {Anderson}},\ }\href {https://doi.org/10.1007/978-3-319-58848-3} {\emph {\bibinfo {title} {The problem of time}}}\ (\bibinfo  {publisher} {Springer Cham},\ \bibinfo {year} {2017})\BibitemShut {NoStop}%
\bibitem [{\citenamefont {Steane}\ \emph {et~al.}(1995)\citenamefont {Steane}, \citenamefont {Szriftgiser}, \citenamefont {Desbiolles},\ and\ \citenamefont {Dalibard}}]{Dalibard95}%
  \BibitemOpen
  \bibfield  {author} {\bibinfo {author} {\bibfnamefont {A.}~\bibnamefont {Steane}}, \bibinfo {author} {\bibfnamefont {P.}~\bibnamefont {Szriftgiser}}, \bibinfo {author} {\bibfnamefont {P.}~\bibnamefont {Desbiolles}},\ and\ \bibinfo {author} {\bibfnamefont {J.}~\bibnamefont {Dalibard}},\ }\bibfield  {title} {\bibinfo {title} {Phase modulation of atomic de broglie waves},\ }\href {https://doi.org/10.1103/PhysRevLett.74.4972} {\bibfield  {journal} {\bibinfo  {journal} {Phys. Rev. Lett.}\ }\textbf {\bibinfo {volume} {74}},\ \bibinfo {pages} {4972} (\bibinfo {year} {1995})}\BibitemShut {NoStop}%
\bibitem [{\citenamefont {Szriftgiser}\ \emph {et~al.}(1996)\citenamefont {Szriftgiser}, \citenamefont {Gu\'ery-Odelin}, \citenamefont {Arndt},\ and\ \citenamefont {Dalibard}}]{Dalibard96}%
  \BibitemOpen
  \bibfield  {author} {\bibinfo {author} {\bibfnamefont {P.}~\bibnamefont {Szriftgiser}}, \bibinfo {author} {\bibfnamefont {D.}~\bibnamefont {Gu\'ery-Odelin}}, \bibinfo {author} {\bibfnamefont {M.}~\bibnamefont {Arndt}},\ and\ \bibinfo {author} {\bibfnamefont {J.}~\bibnamefont {Dalibard}},\ }\bibfield  {title} {\bibinfo {title} {Atomic wave diffraction and interference using temporal slits},\ }\href {https://doi.org/10.1103/PhysRevLett.77.4} {\bibfield  {journal} {\bibinfo  {journal} {Phys. Rev. Lett.}\ }\textbf {\bibinfo {volume} {77}},\ \bibinfo {pages} {4} (\bibinfo {year} {1996})}\BibitemShut {NoStop}%
\bibitem [{\citenamefont {Arndt}\ \emph {et~al.}(1996)\citenamefont {Arndt}, \citenamefont {Szriftgiser}, \citenamefont {Dalibard},\ and\ \citenamefont {Steane}}]{Dalibard96_2}%
  \BibitemOpen
  \bibfield  {author} {\bibinfo {author} {\bibfnamefont {M.}~\bibnamefont {Arndt}}, \bibinfo {author} {\bibfnamefont {P.}~\bibnamefont {Szriftgiser}}, \bibinfo {author} {\bibfnamefont {J.}~\bibnamefont {Dalibard}},\ and\ \bibinfo {author} {\bibfnamefont {A.~M.}\ \bibnamefont {Steane}},\ }\bibfield  {title} {\bibinfo {title} {Atom optics in the time domain},\ }\href {https://doi.org/10.1103/PhysRevA.53.3369} {\bibfield  {journal} {\bibinfo  {journal} {Phys. Rev. A}\ }\textbf {\bibinfo {volume} {53}},\ \bibinfo {pages} {3369} (\bibinfo {year} {1996})}\BibitemShut {NoStop}%
\bibitem [{\citenamefont {Altschul}\ and\ \citenamefont {et~al.}(2015{\natexlab{a}})}]{Microscope17}%
  \BibitemOpen
  \bibfield  {author} {\bibinfo {author} {\bibfnamefont {B.}~\bibnamefont {Altschul}}\ and\ \bibinfo {author} {\bibnamefont {et~al.}},\ }\bibfield  {title} {\bibinfo {title} {Quantum tests of the einstein equivalence principle with the ste–quest space mission},\ }\href {https://doi.org/https://doi.org/10.1016/j.asr.2014.07.014} {\bibfield  {journal} {\bibinfo  {journal} {Advances in Space Research}\ }\textbf {\bibinfo {volume} {55}},\ \bibinfo {pages} {501} (\bibinfo {year} {2015}{\natexlab{a}})}\BibitemShut {NoStop}%
\bibitem [{\citenamefont {Touboul}\ and\ \citenamefont {et~al.}(2022)}]{Microscope22}%
  \BibitemOpen
  \bibfield  {author} {\bibinfo {author} {\bibfnamefont {P.}~\bibnamefont {Touboul}}\ and\ \bibinfo {author} {\bibnamefont {et~al.}} (\bibinfo {collaboration} {MICROSCOPE Collaboration}),\ }\bibfield  {title} {\bibinfo {title} {$microscope$ mission: Final results of the test of the equivalence principle},\ }\href {https://doi.org/10.1103/PhysRevLett.129.121102} {\bibfield  {journal} {\bibinfo  {journal} {Phys. Rev. Lett.}\ }\textbf {\bibinfo {volume} {129}},\ \bibinfo {pages} {121102} (\bibinfo {year} {2022})}\BibitemShut {NoStop}%
\bibitem [{\citenamefont {Armano}\ and\ \citenamefont {et~al.}(2018)}]{LISA18}%
  \BibitemOpen
  \bibfield  {author} {\bibinfo {author} {\bibfnamefont {M.}~\bibnamefont {Armano}}\ and\ \bibinfo {author} {\bibnamefont {et~al.}},\ }\bibfield  {title} {\bibinfo {title} {Beyond the required lisa free-fall performance: New lisa pathfinder results down to $20\text{ }\text{ }\ensuremath{\mu}\mathrm{Hz}$},\ }\href {https://doi.org/10.1103/PhysRevLett.120.061101} {\bibfield  {journal} {\bibinfo  {journal} {Phys. Rev. Lett.}\ }\textbf {\bibinfo {volume} {120}},\ \bibinfo {pages} {061101} (\bibinfo {year} {2018})}\BibitemShut {NoStop}%
\bibitem [{\citenamefont {Armano}\ and\ \citenamefont {et~al.}(2019)}]{LISA19}%
  \BibitemOpen
  \bibfield  {author} {\bibinfo {author} {\bibfnamefont {M.}~\bibnamefont {Armano}}\ and\ \bibinfo {author} {\bibnamefont {et~al.}},\ }\bibfield  {title} {\bibinfo {title} {Lisa pathfinder performance confirmed in an open-loop configuration: Results from the free-fall actuation mode},\ }\href {https://doi.org/10.1103/PhysRevLett.123.111101} {\bibfield  {journal} {\bibinfo  {journal} {Phys. Rev. Lett.}\ }\textbf {\bibinfo {volume} {123}},\ \bibinfo {pages} {111101} (\bibinfo {year} {2019})}\BibitemShut {NoStop}%
\bibitem [{\citenamefont {Schlippert}\ \emph {et~al.}(2014)\citenamefont {Schlippert}, \citenamefont {Hartwig}, \citenamefont {Albers}, \citenamefont {Richardson}, \citenamefont {Schubert}, \citenamefont {Roura}, \citenamefont {Schleich}, \citenamefont {Ertmer},\ and\ \citenamefont {Rasel}}]{QuantumTest14}%
  \BibitemOpen
  \bibfield  {author} {\bibinfo {author} {\bibfnamefont {D.}~\bibnamefont {Schlippert}}, \bibinfo {author} {\bibfnamefont {J.}~\bibnamefont {Hartwig}}, \bibinfo {author} {\bibfnamefont {H.}~\bibnamefont {Albers}}, \bibinfo {author} {\bibfnamefont {L.~L.}\ \bibnamefont {Richardson}}, \bibinfo {author} {\bibfnamefont {C.}~\bibnamefont {Schubert}}, \bibinfo {author} {\bibfnamefont {A.}~\bibnamefont {Roura}}, \bibinfo {author} {\bibfnamefont {W.~P.}\ \bibnamefont {Schleich}}, \bibinfo {author} {\bibfnamefont {W.}~\bibnamefont {Ertmer}},\ and\ \bibinfo {author} {\bibfnamefont {E.~M.}\ \bibnamefont {Rasel}},\ }\bibfield  {title} {\bibinfo {title} {Quantum test of the universality of free fall},\ }\href {https://doi.org/10.1103/PhysRevLett.112.203002} {\bibfield  {journal} {\bibinfo  {journal} {Phys. Rev. Lett.}\ }\textbf {\bibinfo {volume} {112}},\ \bibinfo {pages} {203002} (\bibinfo {year} {2014})}\BibitemShut {NoStop}%
\bibitem [{\citenamefont {van Zoest}\ and\ \citenamefont {et~al.}(2010)}]{MicrogravityEarth10}%
  \BibitemOpen
  \bibfield  {author} {\bibinfo {author} {\bibfnamefont {T.}~\bibnamefont {van Zoest}}\ and\ \bibinfo {author} {\bibnamefont {et~al.}},\ }\bibfield  {title} {\bibinfo {title} {Bose-einstein condensation in microgravity},\ }\href {https://doi.org/10.1126/science.1189164} {\bibfield  {journal} {\bibinfo  {journal} {Science}\ }\textbf {\bibinfo {volume} {328}},\ \bibinfo {pages} {1540} (\bibinfo {year} {2010})},\ \Eprint {https://arxiv.org/abs/https://www.science.org/doi/pdf/10.1126/science.1189164} {https://www.science.org/doi/pdf/10.1126/science.1189164} \BibitemShut {NoStop}%
\bibitem [{\citenamefont {M\"untinga}\ and\ \citenamefont {et~al.}(2013)}]{MicrogravityEarth13}%
  \BibitemOpen
  \bibfield  {author} {\bibinfo {author} {\bibfnamefont {H.}~\bibnamefont {M\"untinga}}\ and\ \bibinfo {author} {\bibnamefont {et~al.}},\ }\bibfield  {title} {\bibinfo {title} {Interferometry with bose-einstein condensates in microgravity},\ }\href {https://doi.org/10.1103/PhysRevLett.110.093602} {\bibfield  {journal} {\bibinfo  {journal} {Phys. Rev. Lett.}\ }\textbf {\bibinfo {volume} {110}},\ \bibinfo {pages} {093602} (\bibinfo {year} {2013})}\BibitemShut {NoStop}%
\bibitem [{\citenamefont {Becker}\ and\ \citenamefont {et~al.}(2018)}]{MAIUS18}%
  \BibitemOpen
  \bibfield  {author} {\bibinfo {author} {\bibfnamefont {D.}~\bibnamefont {Becker}}\ and\ \bibinfo {author} {\bibnamefont {et~al.}},\ }\bibfield  {title} {\bibinfo {title} {Space-borne bose--einstein condensation for precision interferometry},\ }\href {https://doi.org/10.1038/s41586-018-0605-1} {\bibfield  {journal} {\bibinfo  {journal} {Nature}\ }\textbf {\bibinfo {volume} {562}},\ \bibinfo {pages} {391} (\bibinfo {year} {2018})}\BibitemShut {NoStop}%
\bibitem [{\citenamefont {Frye}\ and\ \citenamefont {et~al.}(2021)}]{CAL21}%
  \BibitemOpen
  \bibfield  {author} {\bibinfo {author} {\bibfnamefont {K.}~\bibnamefont {Frye}}\ and\ \bibinfo {author} {\bibnamefont {et~al.}},\ }\bibfield  {title} {\bibinfo {title} {The bose-einstein condensate and cold atom laboratory},\ }\href {https://doi.org/10.1140/epjqt/s40507-020-00090-8} {\bibfield  {journal} {\bibinfo  {journal} {EPJ Quantum Technology}\ }\textbf {\bibinfo {volume} {8}},\ \bibinfo {pages} {1} (\bibinfo {year} {2021})}\BibitemShut {NoStop}%
\bibitem [{\citenamefont {Gaaloul}\ and\ \citenamefont {et~al.}(2022)}]{CAL22}%
  \BibitemOpen
  \bibfield  {author} {\bibinfo {author} {\bibfnamefont {N.}~\bibnamefont {Gaaloul}}\ and\ \bibinfo {author} {\bibnamefont {et~al.}},\ }\bibfield  {title} {\bibinfo {title} {A space-based quantum gas laboratory at picokelvin energy scales},\ }\href {https://doi.org/https://doi.org/10.1038/s41467-022-35274-6} {\bibfield  {journal} {\bibinfo  {journal} {Nat Comm}\ }\textbf {\bibinfo {volume} {13}},\ \bibinfo {pages} {7889} (\bibinfo {year} {2022})}\BibitemShut {NoStop}%
\bibitem [{\citenamefont {Elliott}\ and\ \citenamefont {et~al.}(2023)}]{CAL23}%
  \BibitemOpen
  \bibfield  {author} {\bibinfo {author} {\bibfnamefont {E.~R.}\ \bibnamefont {Elliott}}\ and\ \bibinfo {author} {\bibnamefont {et~al.}},\ }\bibfield  {title} {\bibinfo {title} {Quantum gas mixtures and dual-species atom interferometry in space},\ }\href {https://doi.org/10.1038/s41586-023-06645-w} {\bibfield  {journal} {\bibinfo  {journal} {Nature}\ }\textbf {\bibinfo {volume} {623}},\ \bibinfo {pages} {502} (\bibinfo {year} {2023})}\BibitemShut {NoStop}%
\bibitem [{\citenamefont {Dufour}\ \emph {et~al.}(2014)\citenamefont {Dufour}, \citenamefont {Debu}, \citenamefont {Lambrecht}, \citenamefont {Nesvizhevsky}, \citenamefont {Reynaud},\ and\ \citenamefont {Voronin}}]{GBAR14}%
  \BibitemOpen
  \bibfield  {author} {\bibinfo {author} {\bibfnamefont {G.}~\bibnamefont {Dufour}}, \bibinfo {author} {\bibfnamefont {P.}~\bibnamefont {Debu}}, \bibinfo {author} {\bibfnamefont {A.}~\bibnamefont {Lambrecht}}, \bibinfo {author} {\bibfnamefont {V.~V.}\ \bibnamefont {Nesvizhevsky}}, \bibinfo {author} {\bibfnamefont {S.}~\bibnamefont {Reynaud}},\ and\ \bibinfo {author} {\bibfnamefont {A.~Y.}\ \bibnamefont {Voronin}},\ }\bibfield  {title} {\bibinfo {title} {Shaping the distribution of vertical velocities of antihydrogen in gbar},\ }\href@noop {} {\bibfield  {journal} {\bibinfo  {journal} {The European Physical Journal C}\ }\textbf {\bibinfo {volume} {74}},\ \bibinfo {pages} {2731} (\bibinfo {year} {2014})}\BibitemShut {NoStop}%
\bibitem [{\citenamefont {Cr\'epin}\ \emph {et~al.}(2019)\citenamefont {Cr\'epin}, \citenamefont {Christen}, \citenamefont {Gu\'erout}, \citenamefont {Nesvizhevsky}, \citenamefont {Voronin},\ and\ \citenamefont {Reynaud}}]{GBAR19}%
  \BibitemOpen
  \bibfield  {author} {\bibinfo {author} {\bibfnamefont {P.-P.}\ \bibnamefont {Cr\'epin}}, \bibinfo {author} {\bibfnamefont {C.}~\bibnamefont {Christen}}, \bibinfo {author} {\bibfnamefont {R.}~\bibnamefont {Gu\'erout}}, \bibinfo {author} {\bibfnamefont {V.~V.}\ \bibnamefont {Nesvizhevsky}}, \bibinfo {author} {\bibfnamefont {A.}~\bibnamefont {Voronin}},\ and\ \bibinfo {author} {\bibfnamefont {S.}~\bibnamefont {Reynaud}},\ }\bibfield  {title} {\bibinfo {title} {Quantum interference test of the equivalence principle on antihydrogen},\ }\href {https://doi.org/10.1103/PhysRevA.99.042119} {\bibfield  {journal} {\bibinfo  {journal} {Phys. Rev. A}\ }\textbf {\bibinfo {volume} {99}},\ \bibinfo {pages} {042119} (\bibinfo {year} {2019})}\BibitemShut {NoStop}%
\bibitem [{\citenamefont {Rousselle}\ \emph {et~al.}(2022{\natexlab{a}})\citenamefont {Rousselle}, \citenamefont {Cladé}, \citenamefont {Guellati-Khelifa}, \citenamefont {Guérout},\ and\ \citenamefont {Reynaud}}]{GBAR22}%
  \BibitemOpen
  \bibfield  {author} {\bibinfo {author} {\bibfnamefont {O.}~\bibnamefont {Rousselle}}, \bibinfo {author} {\bibfnamefont {P.}~\bibnamefont {Cladé}}, \bibinfo {author} {\bibfnamefont {S.}~\bibnamefont {Guellati-Khelifa}}, \bibinfo {author} {\bibfnamefont {R.}~\bibnamefont {Guérout}},\ and\ \bibinfo {author} {\bibfnamefont {S.}~\bibnamefont {Reynaud}},\ }\bibfield  {title} {\bibinfo {title} {Analysis of the timing of freely falling antihydrogen},\ }\href {https://doi.org/10.1088/1367-2630/ac5b57} {\bibfield  {journal} {\bibinfo  {journal} {New Journal of Physics}\ }\textbf {\bibinfo {volume} {24}},\ \bibinfo {pages} {033045} (\bibinfo {year} {2022}{\natexlab{a}})}\BibitemShut {NoStop}%
\bibitem [{\citenamefont {Rousselle}\ \emph {et~al.}(2022{\natexlab{b}})\citenamefont {Rousselle}, \citenamefont {Clad{\'e}}, \citenamefont {Guellati-Kh{\'e}lifa}, \citenamefont {Gu{\'e}rout},\ and\ \citenamefont {Reynaud}}]{GBAR22_2}%
  \BibitemOpen
  \bibfield  {author} {\bibinfo {author} {\bibfnamefont {O.}~\bibnamefont {Rousselle}}, \bibinfo {author} {\bibfnamefont {P.}~\bibnamefont {Clad{\'e}}}, \bibinfo {author} {\bibfnamefont {S.}~\bibnamefont {Guellati-Kh{\'e}lifa}}, \bibinfo {author} {\bibfnamefont {R.}~\bibnamefont {Gu{\'e}rout}},\ and\ \bibinfo {author} {\bibfnamefont {S.}~\bibnamefont {Reynaud}},\ }\bibfield  {title} {\bibinfo {title} {Quantum interference measurement of the free fall of anti-hydrogen},\ }\href {https://doi.org/10.1140/epjd/s10053-022-00526-z} {\bibfield  {journal} {\bibinfo  {journal} {The European Physical Journal D}\ }\textbf {\bibinfo {volume} {76}},\ \bibinfo {pages} {209} (\bibinfo {year} {2022}{\natexlab{b}})}\BibitemShut {NoStop}%
\bibitem [{\citenamefont {Altschul}\ and\ \citenamefont {et~al.}(2015{\natexlab{b}})}]{Altschul15}%
  \BibitemOpen
  \bibfield  {author} {\bibinfo {author} {\bibfnamefont {B.}~\bibnamefont {Altschul}}\ and\ \bibinfo {author} {\bibnamefont {et~al.}},\ }\bibfield  {title} {\bibinfo {title} {Quantum tests of the einstein equivalence principle with the ste–quest space mission},\ }\href {https://doi.org/https://doi.org/10.1016/j.asr.2014.07.014} {\bibfield  {journal} {\bibinfo  {journal} {Advances in Space Research}\ }\textbf {\bibinfo {volume} {55}},\ \bibinfo {pages} {501} (\bibinfo {year} {2015}{\natexlab{b}})}\BibitemShut {NoStop}%
\bibitem [{\citenamefont {Battelier}\ and\ \citenamefont {et~al.}(2021)}]{ExpReview21}%
  \BibitemOpen
  \bibfield  {author} {\bibinfo {author} {\bibfnamefont {B.}~\bibnamefont {Battelier}}\ and\ \bibinfo {author} {\bibnamefont {et~al.}},\ }\bibfield  {title} {\bibinfo {title} {Exploring the foundations of the physical universe with space tests of the equivalence principle},\ }\href {https://doi.org/10.1007/s10686-021-09718-8} {\bibfield  {journal} {\bibinfo  {journal} {Experimental Astronomy}\ }\textbf {\bibinfo {volume} {51}},\ \bibinfo {pages} {1695} (\bibinfo {year} {2021})}\BibitemShut {NoStop}%
\bibitem [{\citenamefont {Beau}\ and\ \citenamefont {Martellini}(2024)}]{Beau24}%
  \BibitemOpen
  \bibfield  {author} {\bibinfo {author} {\bibfnamefont {M.}~\bibnamefont {Beau}}\ and\ \bibinfo {author} {\bibfnamefont {L.}~\bibnamefont {Martellini}},\ }\bibfield  {title} {\bibinfo {title} {Quantum delay in the time of arrival of free-falling atoms},\ }\href {https://doi.org/10.1103/PhysRevA.109.012216} {\bibfield  {journal} {\bibinfo  {journal} {Phys. Rev. A}\ }\textbf {\bibinfo {volume} {109}},\ \bibinfo {pages} {012216} (\bibinfo {year} {2024})}\BibitemShut {NoStop}%
\bibitem [{\citenamefont {Davies}(1969)}]{davies1969quantum}%
  \BibitemOpen
  \bibfield  {author} {\bibinfo {author} {\bibfnamefont {E.~B.}\ \bibnamefont {Davies}},\ }\bibfield  {title} {\bibinfo {title} {Quantum stochastic processes},\ }\href@noop {} {\bibfield  {journal} {\bibinfo  {journal} {Communications in Mathematical Physics}\ }\textbf {\bibinfo {volume} {15}},\ \bibinfo {pages} {277} (\bibinfo {year} {1969})}\BibitemShut {NoStop}%
\bibitem [{\citenamefont {Davies}(1970)}]{davies1970repeated}%
  \BibitemOpen
  \bibfield  {author} {\bibinfo {author} {\bibfnamefont {E.~B.}\ \bibnamefont {Davies}},\ }\bibfield  {title} {\bibinfo {title} {On the repeated measurement of continuous observables in quantum mechanics},\ }\href@noop {} {\bibfield  {journal} {\bibinfo  {journal} {Journal of Functional Analysis}\ }\textbf {\bibinfo {volume} {6}},\ \bibinfo {pages} {318} (\bibinfo {year} {1970})}\BibitemShut {NoStop}%
\bibitem [{\citenamefont {Bohr}(1928)}]{Bohr28}%
  \BibitemOpen
  \bibfield  {author} {\bibinfo {author} {\bibfnamefont {N.}~\bibnamefont {Bohr}},\ }\bibfield  {title} {\bibinfo {title} {The quantum postulate and the recent development of atomic theory1},\ }\href {https://doi.org/10.1038/121580a0} {\bibfield  {journal} {\bibinfo  {journal} {Nature}\ }\textbf {\bibinfo {volume} {121}},\ \bibinfo {pages} {580} (\bibinfo {year} {1928})}\BibitemShut {NoStop}%
\bibitem [{\citenamefont {Murdoch}(1987)}]{Murdoch87}%
  \BibitemOpen
  \bibfield  {author} {\bibinfo {author} {\bibfnamefont {D.~R.}\ \bibnamefont {Murdoch}},\ }\href {https://doi.org/10.1017/CBO9780511564307} {\emph {\bibinfo {title} {Niels Bohr's Philosophy of Physics}}}\ (\bibinfo  {publisher} {Cambridge University Press},\ \bibinfo {year} {1987})\BibitemShut {NoStop}%
\bibitem [{\citenamefont {Kleber}(1994)}]{Klebert73}%
  \BibitemOpen
  \bibfield  {author} {\bibinfo {author} {\bibfnamefont {M.}~\bibnamefont {Kleber}},\ }\bibfield  {title} {\bibinfo {title} {Exact solutions for time-dependent phenomena in quantum mechanics},\ }\href {https://doi.org/https://doi.org/10.1016/0370-1573(94)90029-9} {\bibfield  {journal} {\bibinfo  {journal} {Physics Reports}\ }\textbf {\bibinfo {volume} {236}},\ \bibinfo {pages} {331} (\bibinfo {year} {1994})}\BibitemShut {NoStop}%
\bibitem [{SM()}]{SM}%
  \BibitemOpen
  \href@noop {} {}\bibinfo {howpublished} {See Supplementary Material.}\BibitemShut {Stop}%
\bibitem [{\citenamefont {Nesvizhevsky}\ and\ \citenamefont {Voronin}(2015)}]{Nesvizhevsky15}%
  \BibitemOpen
  \bibfield  {author} {\bibinfo {author} {\bibfnamefont {V.~V.}\ \bibnamefont {Nesvizhevsky}}\ and\ \bibinfo {author} {\bibfnamefont {A.~Y.}\ \bibnamefont {Voronin}},\ }\href@noop {} {\emph {\bibinfo {title} {Surprising Quantum Bounces}}}\ (\bibinfo  {publisher} {Imperial College Press, London},\ \bibinfo {year} {2015})\BibitemShut {NoStop}%
\bibitem [{\citenamefont {Busch}(2007)}]{Busch2008}%
  \BibitemOpen
  \bibfield  {author} {\bibinfo {author} {\bibfnamefont {P.}~\bibnamefont {Busch}},\ }\bibinfo {title} {The time--energy uncertainty relation},\ in\ \href {https://doi.org/10.1007/978-3-540-73473-4_3} {\emph {\bibinfo {booktitle} {Time in Quantum Mechanics}}},\ \bibinfo {series and number} {Lecture Notes in Physics},\ \bibinfo {editor} {edited by\ \bibinfo {editor} {\bibfnamefont {J.}~\bibnamefont {Muga}}, \bibinfo {editor} {\bibfnamefont {R.~S.}\ \bibnamefont {Mayato}},\ and\ \bibinfo {editor} {\bibfnamefont {{\'I}.}~\bibnamefont {Egusquiza}}}\ (\bibinfo  {publisher} {Springer Berlin, Heidelberg},\ \bibinfo {year} {2007})\ pp.\ \bibinfo {pages} {73--105}\BibitemShut {NoStop}%
\end{thebibliography}%

\newpage
\onecolumngrid

\vspace{5mm} 

\begin{center}
\textbf{\large SUPPLEMENTARY MATERIAL}
\end{center}

\vspace{5mm} 

\section*{Mean value and standard deviation in the near-field regime}

We start with equation (7) in the main body of the article:
$$
T_x \approx t_c \sqrt{1-\frac{\sigma}{x}\xi},\ \xi\leq \frac{x}{\sigma} ,
$$
where $\frac{\sigma}{x}\gg \max(q,q^2)$. We must have $\frac{\sigma}{x} \gg 1$ or else $T_x \approx t_c$, which, in this case, is equivalent to the semi-classical case that we already know. In this regime, $\frac{x}{\sigma} \ll 1$, and thus, we consider only the region where $\xi \leq 0$. Therefore, we have
$$
T_x \approx t_c \sqrt{\frac{\sigma}{x}}\sqrt{|\xi|},\ \xi\leq 0 ,
$$

Let us recall that the mean value of the stochastic variable $T_x$ is given by
$$
\mathbb{E}(T_x) = \frac{1}{\mathcal{N}}\int_{-\infty}^{0}T_x(\xi) \frac{1}{\sqrt{2\pi}}e^{-\frac{\xi^2}{2}} ,
$$
where 
$$
\mathcal{N} = \int_{-\infty}^{0}\frac{1}{\sqrt{2\pi}}e^{-\frac{\xi^2}{2}} = \frac{1}{2}
$$
is the normalization factor of the distribution. 

Hence, we find:
$$
\mathbb{E}(T_x) = 2t_c \sqrt{\frac{\sigma}{x}}\int_{-\infty}^{0}\sqrt{|\xi|} \frac{1}{\sqrt{2\pi}}e^{-\frac{\xi^2}{2}} = \frac{2^{1/4}}{\sqrt{\pi}}\Gamma(3/4)\  t_c\sqrt{\frac{\sigma}{x}} .
$$

As the mean value of the stochastic variable $T_x^2$ is given by
$$
\mathbb{E}(T_x^2) = \frac{1}{\mathcal{N}}\int_{-\infty}^{0}T_x(\xi)^2 \frac{1}{\sqrt{2\pi}}e^{-\frac{\xi^2}{2}} ,
$$
we obtain
$$
\mathbb{E}(T_x^2) = 2t_c^2 \frac{\sigma}{x}\int_{-\infty}^{0}|\xi| \frac{1}{\sqrt{2\pi}}e^{-\frac{\xi^2}{2}} = \sqrt{\frac{2}{\pi}}\ \frac{\sigma}{x}t_c^2.
$$

\section*{Uncertainty for the energy of a free-falling particle}

The Hamiltonian $\hat{H}$ for a free-falling quantum particle is given by:
$$
    \hat{H} = \frac{\hat{p}^2}{2m}-mg\hat{x} ,
$$
where $\hat{p} = -i\hbar\frac{\partial}{\partial x}$ is the momentum operator and $\hat{x}$ is the position operator.  
Consider the initial Gaussian wavepacket:
\begin{equation}\label{Eq:InitialWavePacket}
    \psi_0(x) = \frac{1}{(2\pi\sigma^2)^{1/4}}e^{-\frac{x^2}{4\sigma^2}}\ ,
\end{equation}
where the mean value of the particle's momentum is zero $\langle \hat{p}\rangle = 0$, and where $\sigma$ is the standard deviation of the position operator $\sqrt{\langle \hat{x}^2\rangle-\langle \hat{x}\rangle^2}=\sigma$, with $\langle \hat{x}\rangle = 0$. 
The initial Gaussian wavepacket can also be expressed in the momentum basis through the Fourier transform of \eqref{Eq:InitialWavePacket}
$$
    \psi_0(p) = \frac{1}{(2\pi \sigma_p^2)^{1/4}}e^{-\frac{p^2}{4\sigma_p^2} }\ ,
$$
where the standard deviation of the momentum operator is $\sqrt{\langle \hat{p}^2 \rangle - \langle \hat{p} \rangle^2} = \sigma_p = \frac{\hbar}{2\sigma}$. 

We may calculate the mean value $\langle \hat{H} \rangle$  of the energy operator as:
\begin{equation}\label{Eq:MeanEnergy}
\langle \hat{H} \rangle  =  \langle \frac{\hat{p}^2}{2m} \rangle - \langle mgx \rangle = \langle \frac{\hat{p}^2}{2m} \rangle + 0 
= \frac{\sigma_p^2}{2m} =  \frac{\hbar^2}{8m\sigma^2} .
\end{equation}
We also calculate the mean squared value of the energy operator as:
\begin{equation}\label{Eq:MeanEnergySquared}
\langle \hat{H}^2 \rangle  = \langle \frac{\hat{p}^4}{4m^2} \rangle + \langle m^2g^2\hat{x}^2 \rangle - mg\langle (\hat{p}^2\hat{x}+\hat{x}\hat{p}^2 )\rangle = \frac{3}{4m^2}\sigma_p^4 + m^2g^2\sigma^2 = \frac{3\hbar^4}{64m^2\sigma^4} + m^2g^2\sigma^2  ,
\end{equation}
where we used $\langle \hat{x}^2\rangle = \sigma^2$, $\langle \hat{p}^4\rangle = 3\sigma_p^4$, and $\langle (\hat{p}^2\hat{x}+\hat{x}\hat{p}^2 )\rangle =0$, as can be seen from:
\begin{align*}
\langle (\hat{p}^2\hat{x}+\hat{x}\hat{p}^2 )\rangle &= \langle (2\hat{x}\hat{p}^2-2i\hbar \hat{p})\rangle \\
&= \langle 2\hat{x}\hat{p}^2)\rangle +0  \\
&= -2\hbar^2\int_{-\infty}^{+\infty} dx\ x \psi_0(x)^\ast\frac{\partial^2}{\partial x^2}\psi_0(x) \\
&= -2\hbar^2\int_{-\infty}^{+\infty} dx\ x \psi_0(x)^\ast\frac{\partial}{\partial x}\left(-\frac{x}{2\sigma^2}\psi_0(x)\right) \\
&= -2\hbar^2\int_{-\infty}^{+\infty} dx\ x \psi_0(x)^\ast\left(-\frac{1}{2\sigma^2}+\frac{x^2}{4\sigma^4}\right)\psi_0(x) \\
&= -2\hbar^2\int_{-\infty}^{+\infty} dx\ \left(-\frac{x}{2\sigma^2}+\frac{x^3}{4\sigma^4}\right)|\psi_0(x)|^2 \\
&= -2\hbar^2\left(-\frac{\langle x\rangle}{2\sigma^2}+\frac{\langle x^3\rangle}{4\sigma^4}\right),
\end{align*}
and the announced result follows from since $\langle x\rangle = 0$ and $\langle x^3\rangle = 0$. Combining \eqref{Eq:MeanEnergy} and \eqref{Eq:MeanEnergySquared}, we find that the standard deviation of the energy operator is given as:
\begin{equation}
    \Delta E^2 \equiv \langle \hat{H}^2\rangle - \langle \hat{H}\rangle^2 = \frac{3\hbar^4}{64m^2\sigma^4} + m^2g^2\sigma^2 - \frac{\hbar^4}{64m^2\sigma^4} = \frac{\hbar^4}{32m^2\sigma^4} + m^2g^2\sigma^2 =  m^2g^2\sigma^2\left(1+\frac{\hbar^4}{32m^4g^2\sigma^6}\right) ,
\end{equation}
from which we finally obtain:
\begin{equation}
    \Delta E = mg\sigma\sqrt{1+\frac{\hbar^4}{32m^4g^2\sigma^6}} .
\end{equation}

\end{document}